\documentclass[12pt,letterpaper]{article}
\usepackage{amsmath,amssymb,array,calc,rotating,epsfig,psfrag,amscd, cite}

\numberwithin{equation}{section}

\makeatletter
\renewcommand\section{\@startsection {section}{1}{\z@}%
                                   {-3.5ex \@plus -1ex \@minus -.2ex}
                                   {2.3ex \@plus.2ex}%
                                   {\normalfont\large\bfseries}}
\renewcommand\subsection{\@startsection{subsection}{2}{\z@}%
                                     {-3.25ex\@plus -1ex \@minus -.2ex}%
                                     {1.5ex \@plus .2ex}%
                                     {\normalfont\bfseries}}
\makeatother

\let\non\nonumber

\let\s=\sigma

\let\S=\Sigma

\newcommand{\bea}{\begin{eqnarray}}
\newcommand{\eea}{\end{eqnarray}}
\newcommand{\be}{\begin{equation}}
\newcommand{\ee}{\end{equation}}

\newcommand{\p}{\partial}

\newcommand{\C}[1]{$(\ref{#1})$}

\def\IZ{\relax\ifmmode\mathchoice
{\hbox{\cmss Z\kern-.4em Z}}{\hbox{\cmss Z\kern-.4em Z}}
{\lower.9pt\hbox{\cmsss Z\kern-.4em Z}} {\lower1.2pt\hbox{\cmsss
Z\kern-.4em Z}}\else{\cmss Z\kern-.4em Z}\fi}
\def\IR{\relax{\rm I\kern-.18em R}}

\def\one{{\hbox{ 1\kern-.8mm l}}}

\newlength{\bredde}
\def\slash#1{\settowidth{\bredde}{$#1$}\ifmmode\,\raisebox{.15ex}{/}
\hspace*{-\bredde} #1\else$\,\raisebox{.15ex}{/}\hspace*{-\bredde}
#1$\fi}

\newsavebox{\zzzbar}
\sbox{\zzzbar}
  {\setlength{\unitlength}{0.9em}
  \begin{picture}(0.6,0.7)
  \thinlines
  \put(0,0){\line(1,0){0.6}}
  \put(0,0.75){\line(1,0){0.575}}
  \multiput(0,0)(0.0125,0.025){30}{\rule{0.3pt}{0.3pt}}
  \multiput(0.2,0)(0.0125,0.025){30}{\rule{0.3pt}{0.3pt}}
  \put(0,0.75){\line(0,-1){0.15}}
  \put(0.015,0.75){\line(0,-1){0.1}}
  \put(0.03,0.75){\line(0,-1){0.075}}
  \put(0.045,0.75){\line(0,-1){0.05}}
  \put(0.05,0.75){\line(0,-1){0.025}}
  \put(0.6,0){\line(0,1){0.15}}
  \put(0.585,0){\line(0,1){0.1}}
  \put(0.57,0){\line(0,1){0.075}}
  \put(0.555,0){\line(0,1){0.05}}
  \put(0.55,0){\line(0,1){0.025}}
  \end{picture}}

\def\Im{{\rm Im ~}}

\newcommand{\ena}{\end{eqnarray}}
\newcommand{\beqa}{\begin{eqnarray}}
\newcommand{\eeqa}{\end{eqnarray}}

\def\s{\sigma}

\def\S{\Sigma}

\begin{document}
\begin{titlepage}

\begin{center}



\vskip 2 cm
{\Large \bf Poisson equation for genus two string invariants: a conjecture}\\
\vskip 1.25 cm { Anirban Basu\footnote{email address:
    anirbanbasu@hri.res.in} } \\
{\vskip 0.5cm  Harish--Chandra Research Institute, HBNI, Chhatnag Road, Jhusi,\\
Prayagraj 211019, India}

\end{center}

\vskip 2 cm

\begin{abstract}
\baselineskip=18pt

We consider some string invariants at genus two that appear in the analysis of the $D^8\mathcal{R}^4$ and $D^6\mathcal{R}^5$ interactions in type II string theory. We conjecture a Poisson equation involving them and the Kawazumi--Zhang invariant based on their asymptotic expansions around the non--separating node in the moduli space of genus two Riemann surfaces.

\end{abstract}

\end{titlepage}


\section{Introduction}

String invariants or modular graph forms~\cite{DHoker:2015gmr,DHoker:2015wxz} arise as integrands in the integrals over moduli space of Riemann surfaces resulting from the analysis of local interactions in the low momentum expansion of amplitudes in string theory. Hence understanding their detailed properties is central to obtaining the coefficients of various interactions in the effective action. These graphs have links given by the Green function or their worldsheet derivatives, while the vertices are the positions of insertions of vertex operators on the punctured worldsheet. Obtaining eigenvalue equations satisfied by these graphs is very useful in performing the integrals over moduli space, along with a knowledge of their asymptotic expansions around various degenerating nodes of the worldsheet. This has yielded various results at genus one~\cite{Green:1999pv,Green:2008uj,Richards:2008jg,Green:2013bza,DHoker:2015gmr,Basu:2015ayg,DHoker:2015wxz,Zerbini:2015rss,DHoker:2016mwo,Basu:2016xrt,Basu:2016kli,Basu:2016mmk,DHoker:2016quv,Kleinschmidt:2017ege,DHoker:2019blr,Basu:2019idd,Gerken:2019cxz,Gerken:2020yii,Gerken:2020aju,Gerken:2020xfv} and two~\cite{DHoker:2005vch,DHoker:2005jhf,Berkovits:2005df,Berkovits:2005ng,DHoker:2013fcx,DHoker:2014oxd,Pioline:2015qha,DHoker:2017pvk,DHoker:2018mys,Basu:2018bde,DHoker:2020prr,DHoker:2020tcq,Basu:2020pey,DHoker:2020uid,Basu:2020iok,DHoker:2020aex,Basu:2020goe} leading to an intricate underlying structure.      

The modular graph functions that arise at genus one in type II string theory are $SL(2,\mathbb{Z})_\tau$ invariant functions of the complex structure $\tau$ of the torus. These graphs satisfy Poisson equations which have been derived in various cases. It is interesting to analyze the structure of these equations simply based on the Laurent expansion of these graphs around the cusp ${\Im}\tau \rightarrow \infty$. This expansion has several power behaved terms apart from terms that are exponentially suppressed. Now simply based on the structure of the power behaved terms for a given graph, one can try to guess the Poisson equation it satisfies, as was originally done in~\cite{DHoker:2015gmr} hence providing a powerful tool to search for potential eigenvalue equations.  

The analysis of obtaining eigenvalue equations for string invariants gets considerably more involved at genus two. The $Sp(4,\mathbb{Z})$ invariant graph with one link that arises as the integrand over moduli space of the $D^6\mathcal{R}^4$ and the $D^4\mathcal{R}^5$ interactions in the low momentum expansion of the four and five graviton amplitudes respectively~\cite{DHoker:2005vch,DHoker:2020prr,DHoker:2020tcq}, is the Kawazumi--Zhang (KZ) invariant~\cite{Kawazumi,Zhang,DHoker:2013fcx} which satisfies a Poisson equation on moduli space, which has enabled the calculation of the coefficients of the $D^6\mathcal{R}^4$ and $D^4\mathcal{R}^5$ interactions in the effective action~\cite{DHoker:2014oxd}. What about such Poisson equations for graphs with more than one link? Unlike the analysis at genus one, there are no such known equations\footnote{Differential equations involving four and six derivatives on moduli space have been obtained in~\cite{Basu:2020goe}. However, the primary equation leading to these results involves two derivatives.}, though the asymptotic expansion of various graphs with more than one link around the degenerating nodes of the genus two Riemann surface have been analyzed in detail~\cite{DHoker:2017pvk,DHoker:2018mys}. Thus in analogy with the analysis involving their genus one counterparts, it is natural to ask if these asymptotic expansions can be used to guess any eigenvalue equation satisfied by genus two string invariants. 

Based on the asymptotic expansions around the non--separating node of the genus two Riemann surface of several graphs that arise in the analysis of the $D^8\mathcal{R}^4$ and the $D^6\mathcal{R}^5$ interactions, we shall argue that there is a candidate Poisson equation that arises naturally involving these graphs as well as the KZ invariant, which we conjecture to be true over all of moduli space. It will be interesting to check this claim along the lines of~\cite{DHoker:2014oxd,Basu:2018bde}.                                          
  
After briefly reviewing relevant details about genus two Riemann surfaces and in particular the non--separating node, we shall demonstrate how the Poisson equation satisfied by the KZ invariant can be guessed from its asymptotic expansion around the non--separating node, reproducing the known result. We then proceed similarly to guess a Poisson equation involving graphs with more than one link.          

\section{Genus two asymptotics and the Kawazumi--Zhang invariant}

We denote the genus two worldsheet by $\S_2$, and the conformally invariant Arakelov Green function by $G(z,w)$. The imaginary part of the period matrix is defined by ${\rm Im}\Omega = Y$. We define the inverse matrix $Y^{-1}_{IJ} = (Y^{-1})_{IJ}$ ($I,J=1,2$), and the dressing factors
\be (z,\overline{w}) = Y^{-1}_{IJ} \omega_I (z) \overline{\omega_J (w)}, \quad \mu (z) = (z,\overline{z}), \quad P(z,w) = (z,\overline{w})(w,\overline{z})\ee
which we often use, where $\omega_I = \omega_I (z) dz$ is the Abelian differential one form.   
The integration measure over the worldsheet is given by $d^2 z = i dz \wedge d\overline{z}$.  

The asymptotic expansions of the various graphs around the non--separating node shall play a central role in our analysis, which we briefly review. Parametrizing the period matrix $\Omega$ as
\be \label{parap}\Omega= \left( \begin{array}{cc} \tau & v \\ v & \sigma \end{array} \right),\ee  
the non--separating node is obtained by taking $\sigma \rightarrow i\infty$, while keeping $\tau, v$ fixed\footnote{The other contribution coming from taking $\tau \rightarrow i\infty$, while keeping $\sigma, v$ fixed, is simply related to this by $\tau \leftrightarrow \sigma$ exchange, and hence we consider only one of them.}. At this node, an $SL(2,\mathbb{Z})_\tau$ subgroup of $Sp(4,\mathbb{Z})$ survives whose action on $v, \tau$ and $\sigma$ is given by~\cite{Moore:1986rh,DHoker:2017pvk,DHoker:2018mys}
\be v \rightarrow \frac{v}{(c\tau+d)}, \quad \tau \rightarrow \frac{a\tau+b}{c\tau+d},\quad \sigma \rightarrow \sigma - \frac{cv^2}{c\tau+d},\ee 
where $a,b,c,d \in \mathbb{Z}$ and $ad-bc=1$. Here $v$ parametrizes the coordinate on the torus with complex structure $\tau$, and hence
\be -\frac{1}{2} \leq v_1 \leq \frac{1}{2}, \quad 0 \leq v_2 \leq \tau_2.\ee
Also $\sigma_2$ along with $v_2$ and $\tau_2$ forms the $SL(2,\mathbb{Z})_\tau$ invariant quantity
\be \label{deft}t= \sigma_2 - \frac{v_2^2}{\tau_2}\ee
which is a useful parameter in the asymptotic expansion. 

We shall also make use of the $SL(2,\mathbb{Z})_\tau$ invariant operators
\be \Delta_\tau = 4\tau_2^2 \p_\tau \overline\p_\tau , \quad \Delta_v = 4\tau_2 \p_v \overline\p_v \ee
in our analysis.

In order to motivate the Poisson equation, we shall use the expression for the $Sp(4,\mathbb{Z})$ invariant Laplacian $\Delta$ expanded around the non--separating node. This is given by~\cite{Pioline:2015qha}\footnote{We ignore terms involving $\p/\p\sigma_1$ in \C{Laplacian} as they are not relevant to us.} 
\be \label{Laplacian}\Delta = t^2\frac{\p^2}{\p t^2} - t\frac{\p}{\p t} +\frac{t}{2} \Delta_v + \Delta_\tau .\ee 

We now list various $SL(2,\mathbb{Z})_\tau$ covariant expressions on the toroidal worldsheet $\Sigma$ with complex structure $\tau$ and coordinate $v$ that will be relevant for our purposes. They arise in the analysis of the genus two string invariants in their asymptotic expansions around the non--separating node, where the worldsheet is given by $\S$ with two additional punctures (beyond those from the vertex operators) connected by a long, thin handle whose proper length is proportional to $t$, hence providing a physical interpretation for this parameter.

The Green function $g(v)$ at genus one is given by
\be \label{Green}g(v)\equiv g(v;\tau) = \sum_{(m,n) \neq (0,0)}\frac{\tau_2}{\pi\vert m+n\tau\vert^2} e^{2\pi i(my-nx)},\ee
where we have parametrized $v$ as
\be v = x+\tau y,\ee
with $x,y \in (0,1]$. It satisfies the differential equations
\be \label{diff}\Delta_\tau g(v)=0, \quad \Delta_v g(v) = -4\pi \tau_2 \delta^2 (v) +4\pi,\ee
where the delta function is normalized to satisfy $\int_\S d^2 z \delta^2(z)=2$ (note that $d^2 z = i dz \wedge d\overline{z}$ on $\S$).

The iterated Green function $g_{k+1} (v)$ is defined recursively by
\be g_{k+1} (v) \equiv g_{k+1} (v;\tau) =  \int_{\Sigma} \frac{d^2 z}{2\tau_2} g(v-z)g_k (z)\ee
for $k \geq 1$, where $g_1 (v) = g(v)$ by definition. The non--holomorphic Eisenstein series is defined by
\be E_{k+1} \equiv E_{k+1} (\tau) = g_{k+1} (0)\ee
for $k \geq 1$. They satisfy the differential equations
\bea \Delta_\tau g_{k+1} (v) = k(k+1) g_{k+1} (v), \quad \Delta_\tau E_{k+1} = k(k+1) E_{k+1}, \quad \Delta_v g_{k+1} (v) = -4\pi g_k (v).\non \\ \eea 

We next consider the family of elliptic modular graph functions~\cite{DHoker:2018mys} defined by 
\be \label{defF}F_{2k} (v) = \frac{1}{(2k)!} \int_{\S} \frac{d^2 z}{2\tau_2} f(z)^{2k},\ee
where $k \geq 1$ is a positive integer, and $f(z) = g(v-z) -g(z)$.

From the definition and using the results mentioned above, we get that
\be \label{F2}F_2 (v) = E_2 - g_2 (v)\ee
which satisfies the equations
\be \Delta_\tau  F_2 (v) = 2 F_2 (v), \quad \Delta_v F_2 (v) = 4\pi g(v).\ee
Thus we see that both $\Delta_\tau$ and $\Delta_v$ acting on $F_2 (v)$ yield elementary $SL(2,\mathbb{Z})_\tau$ invariant quantities.

The expression for $F_4 (v)$ which directly follows from \C{defF} is not relevant for our purposes. However, an analysis of this expression yields the Poisson equation~\cite{DHoker:2020tcq,Basu:2020pey,DHoker:2020aex}
\be \label{F4}\Big(\Delta_\tau -2\Big) \Big( F_4 (v) - \frac{1}{2}F_2(v)^2 \Big)= -2 F_2 (v)^2\ee
which shall be useful. 
Thus so far as terms involving derivatives on moduli space are concerned, $\Delta_\tau F_4 (v)$ can be expressed in terms of $\Delta_\tau F_2(v)^2$. 
Also note that $\Delta_v F_4 (v)$ does not yield anything particularly useful.   

These relations mentioned above will be repeatedly used in our analysis below.

\begin{figure}[ht]
\begin{center}
\[
\mbox{\begin{picture}(130,45)(0,0)
\includegraphics[scale=.7]{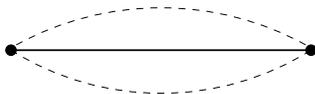}
\end{picture}}
\]
\caption{The string invariant $\varphi$}
\end{center}
\end{figure}

Let us now consider the Kawazumi--Zhang (KZ) invariant which arises in the analysis of the $D^6\mathcal{R}^4$ and the $D^4\mathcal{R}^5$ interactions at genus two. This string invariant is defined by  
\be \label{KZ}\varphi (\Omega,\overline\Omega) = \frac{1}{4}\int_{\S_2^2} \prod_{i=1}^2 d^2 z_i G(z_1,z_2) P(z_1,z_2)\ee
as depicted by figure 1.
Its asymptotic expansion around the non--separating node is given by~\cite{Wentworth,Jong,DHoker:2013fcx,Pioline:2015qha,DHoker:2017pvk,DHoker:2018mys}
\be \label{exp1}\varphi = \frac{\pi t}{6} +\frac{g(v)}{2} +\frac{5F_2 (v)}{4\pi t} +O(e^{-2 \pi t}),\ee
which has only a finite number of power behaved terms in $t$ while the remaining terms are exponentially suppressed. Thus from the expression for the Laplacian in \C{Laplacian}, we get that 
\be \label{exact}\Big(\Delta -5 \Big)\varphi = -\pi {\rm det}Y \delta^2 (v)\ee
up to exponentially suppressed contributions. One might guess that \C{exact} is the exact Poisson equation satisfied by the KZ invariant over all of moduli space and not just in an asymptotic expansion around the non--separating node keeping only the power behaved terms in $t$, which in fact turns out to be correct~\cite{DHoker:2014oxd}\footnote{In fact, the complete asymptotic expansion of $\varphi$, including the exponentially suppressed contributions in \C{exp1} have been obtained in~\cite{Pioline:2015qha} by directly solving \C{exact}.}. Thus we see that the asymptotic expansion provides a good starting point for guessing the exact eigenvalue equation.

As an aside, note that the asymptotic expansion of the various string invariants around the separating node where $v \rightarrow 0$ with $\tau,\s$ finite in \C{parap} is a Taylor series in ${\rm ln} \vert \lambda \vert$ with a finite number of terms, along with an infinite number of corrections that are potentially of the form $\vert \lambda \vert^m ({\rm ln} \vert \lambda \vert)^n$ for $m > 0$, $n \geq 0$ where~\cite{DHoker:2018mys}    
\be \lambda = 2\pi v \eta^2 (\tau) \eta^2 (\s).\ee
Hence this  asymptotic expansion is not particularly useful in trying to derive any Poisson equations as the action of the Laplacian mixes the various contributions.

On the other hand, for the purposes of trying to guess Poisson equations satisfied by the string invariants, their asymptotic expansions around the non--separating node are very useful. This is because such expansions involve only a finite number of terms in the Laurent series expanded around $t \rightarrow \infty$, along with exponentially suppressed contributions~\cite{DHoker:2018mys}. Thus acting with the Laplacian in \C{Laplacian} on the asymptotic series, we can simply focus on the finite number of power behaved terms, as there is no mixing with the exponentially suppressed terms.

\section{A Poisson equation for genus two string invariants}

Based on the discussion above, we now analyze string invariants with more than one link, with the aim of trying to motivate a Poisson equation satisfied by them. We first consider the graphs that arise in the integrand over moduli space of the $D^8\mathcal{R}^4$ interaction in the low momentum expansion of the four graviton amplitude, each of which has two links. Their asymptotic expansions around the non--separating node are given in~\cite{DHoker:2018mys} up to exponentially suppressed contributions. One of these graphs, denoted $\mathcal{Z}_1$, forms a closed loop on the worldsheet and its asymptotic expansion is significantly more complicated than the others, and we do not consider it.  

\begin{figure}[ht]
\begin{center}
\[
\mbox{\begin{picture}(270,130)(0,0)
\includegraphics[scale=.65]{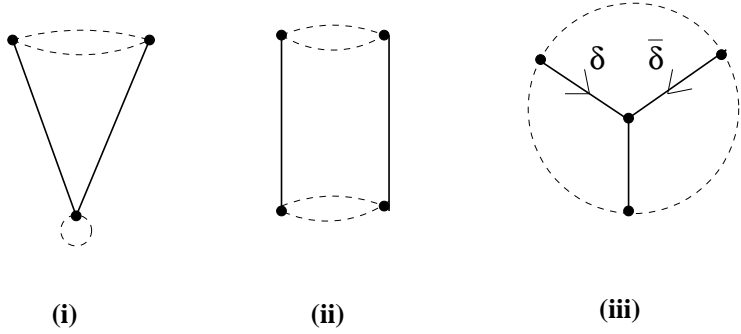}
\end{picture}}
\]
\caption{The string invariants (i) $\mathcal{Z}_2$, (ii) $\mathcal{Z}_3$ and (iii) $\mathcal{Z}_5$}
\end{center}
\end{figure}

We consider the graphs $\mathcal{Z}_2$ and $\mathcal{Z}_3$ defined by
\bea \mathcal{Z}_2 (\Omega,\overline\Omega)&=& -\frac{1}{4}\int_{\S_2^3} \prod_{i=1}^3 d^2 z_i G(z_1,z_2) G(z_1,z_3) \mu (z_1)P(z_2,z_3), \non \\ \mathcal{Z}_3 (\Omega,\overline\Omega)&=&\frac{1}{8}\int_{\S_2^4} \prod_{i=1}^4 d^2 z_i G(z_1,z_4) G(z_2,z_3)P(z_1,z_2)P(z_3,z_4)\eea
as depicted by figure 2.
The asymptotic expansion around the non--separating node of $\mathcal{Z}_2$ is given by
\bea \label{Z2}\mathcal{Z}_2 &=& -\frac{7\pi^2 t^2}{90} -\frac{\pi t}{3}g(v) - E_2 - \frac{g(v)^2}{2} +\frac{F_2 (v)}{2} \non \\ &&+\frac{1}{\pi t}\Big[-2\Big(E_3 -g_3 (v)\Big)+\frac{1}{2} g(v)F_2 (v)  -\frac{1}{16\pi} \Delta_v \Big(F_2(v)^2 + 2 F_4 (v)\Big)\Big] \non \\ &&-\frac{1}{4\pi^2 t^2} \Big[\Delta_\tau +5\Big]F_4 (v) + O(e^{-2\pi t})\eea
where we have kept all the power behaved terms. In \C{Z2} we have used the relation~\cite{DHoker:2015gmr,DHoker:2016mwo,Basu:2016kli}
\be \label{defD}D_3 = E_3 +\zeta(3)\ee
between modular graphs to rewrite the expression in~\cite{DHoker:2018mys}, where the dihedral graph $D_3$ is defined by
\be D_3 \equiv D_3 (\tau) = \int_\S \frac{d^2 z}{2\tau_2} g(z)^3.\ee
On the other hand, the asymptotic expansion of $\mathcal{Z}_3$ is given by
\bea \label{Z3} \mathcal{Z}_3 &=& \frac{\pi^2 t^2}{18} +\frac{\pi t}{3}g(v) + \frac{g(v)^2}{2} +\frac{F_2 (v)}{6} +\frac{1}{\pi t} \Big[ -\frac{1}{2} g(v) F_2 (v) +\frac{1}{8\pi} \Delta_v F_2 (v)^2\Big] \non \\ &&+\frac{1}{8\pi^2 t^2} \Big[ \Delta_\tau + 5\Big] F_2(v)^2 + O(e^{-2\pi t}).\eea
Now our aim is to obtain a Poisson equation involving the minimal number of graphs, which is the simplest possible setting. From \C{Z2} and \C{Z3} we see that apart from differences involving various other contributions, $\mathcal{Z}_2$ contains $F_4 (v)$ in its asymptotic expansion while $\mathcal{Z}_3$ does not\footnote{In fact, $\mathcal{Z}_3$ is completely determined by the KZ invariant~\cite{Basu:2020goe}.}. Thus we immediately see that any Poisson equation (in fact, any equation) involving these two graphs must at least involve another graph which has $F_4 (v)$ in its asymptotic expansion. A natural candidate is the string invariant $\mathcal{Z}_5$ which arises in the analysis of the $D^6\mathcal{R}^5$ interaction in the low momentum expansion of the five graviton amplitude~\cite{DHoker:2020tcq}. It is defined by
\be \mathcal{Z}_5 (\Omega,\overline\Omega)= -\frac{2}{\pi}\int_{\S_2^4} \prod_{i=1}^4 d^2 z_i G(z_1,z_4) \p_{z_1} G(z_1,z_2) \overline\p_{z_1} G(z_1,z_3) (z_2,\overline{z}_4) (z_4,\overline{z}_3)(z_3,\overline{z}_2)\ee   
as depicted by figure 2. Thus it has three links, where two of them are given by the worldsheet (anti)holomorphic derivatives of the Green function (depicted by $\delta$ and $\overline\delta$ in figure 2). Its asymptotic expansion around the non--separating node is given by 
\bea \label{Z5} \mathcal{Z}_5 &=& \frac{3}{2\pi t} \Big[ -8 \Big(D_3 - D_3^{(1)} (v)\Big) + 24 g(v) F_2 (v) +32 \Big(E_3 - g_3 (v)\Big)- \frac{1}{\pi} \Delta_v F_2(v)^2\Big]\non \\ &&+\frac{3}{2\pi^2 t^2} \Big[ 20 F_4 (v)  -\Big(\Delta_\tau -6\Big)F_2(v)^2\Big]+ O(e^{-2\pi t}),\eea
where we have used \C{F4} to write the expression differently compared to~\cite{DHoker:2020tcq}. Also, in \C{Z5} we have defined~\cite{DHoker:2018mys} the elliptic modular graph
\be D_3^{(1)} (v) \equiv D_3^{(1)} (v;\tau) = \int_{\S} \frac{d^2 z}{2\tau_2} g(z-v) g(z)^2\ee
depicted by figure 3. Thus note that $D_3^{(1)} (0) = D_3$.

\begin{figure}[ht]
\begin{center}
\[
\mbox{\begin{picture}(110,50)(0,0)
\includegraphics[scale=.7]{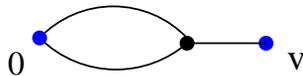}
\end{picture}}
\]
\caption{The elliptic modular graph $D_3^{(1)} (v)$}
\end{center}
\end{figure}

Now in both the asymptotic expansions of $\mathcal{Z}_3$ and $\mathcal{Z}_5$ given by \C{Z3} and \C{Z5} respectively, the only terms involving derivatives on moduli space involve $\Delta_v F_2(v)^2$ and $\Delta_\tau F_2(v)^2$. In fact, the total contribution in either case is proportional to     
\be \label{prop}\frac{\Delta_v F_2(v)^2}{\pi^2 t} + \frac{\Delta_\tau F_2(v)^2}{\pi^2 t^2},\ee
and hence the combination 
\be \label{step1}\mathcal{Z}_5 + 12 \mathcal{Z}_3\ee 
contains no terms with derivatives on moduli space. Furthermore, the $O(t^2)$ and $O(t)$ contributions in \C{step1} matches those in the expansion of $24\varphi^2$ using \C{exp1}. Thus we get that
\bea \label{step2}\mathcal{Z}_5 +12\mathcal{Z}_3 - 24 \varphi^2 &=& -8 F_2 (v) +\frac{12}{\pi t} \Big[ -\Big(D_3 - D_3^{(1)} (v)\Big) + 4\Big(E_3 - g_3 (v)\Big)\Big]\non \\ &&+\frac{3}{\pi^2 t^2} \Big[ 10 F_4 (v) - 7 F_2(v)^2\Big]+ O(e^{-2\pi t}).\eea
Let us compare the terms involving $F_4 (v)$ in \C{Z2} and \C{step2}, even though the other contributions are very different. While \C{step2} has only $F_4 (v)$ in its asymptotic expansion, \C{Z2} has $F_4 (v)$, $\Delta_v F_4 (v)$ and $\Delta_\tau F_4 (v)$. Thus it is natural to ask if the action of the $Sp(4,\mathbb{Z})$ invariant Laplacian on the combination of graphs $\mathcal{Z}_5 +12\mathcal{Z}_3 - 24 \varphi^2$ might be related in any way to $\mathcal{Z}_2$. Hence let us consider the action of \C{Laplacian} on \C{step2}. Apart from its action on $D_3^{(1)} (v)$ in \C{step2}, its action on the other terms can be obtained using the various results given in the previous section and \C{defD}.     

While the action of $\Delta_v$ on $D_3^{(1)} (v)$ is given by
\be \Delta_v D_3^{(1)} (v) = 4\pi\Big(E_2 - g(v)^2\Big),\ee
the action of $\Delta_\tau$ on $D_3^{(1)} (v)$ is given by
\be \label{acttau}\Delta_\tau D_3^{(1)}(v) = 2E_3 + 4 g_3 (v) + 2 g(v) F_2 (v) - \frac{1}{4\pi}\Delta_v F_2(v)^2.\ee
We present the derivation of \C{acttau} in the appendix.

Putting together the various contributions, we obtain the asymptotic expansion
\bea \label{step3}&&\Delta \Big(\mathcal{Z}_5 +12\mathcal{Z}_3 - 24 \varphi^2\Big) = -16\pi t g(v) -16 F_2 (v) + 24 E_2 + 96 g_2 (v) - 24 g(v)^2 \non \\ &&+\frac{3}{\pi t}\Big[ 128\Big(E_3 - g_3 (v)\Big) -12 \Big(D_3 - D_3^{(1)} (v)\Big) + 8 g(v) F_2 (v) +\frac{1}{\pi} \Delta_v \Big( 5\mathcal{F}_4 (v) - \frac{9}{2} F_2(v)^2\Big)\Big]\non \\ &&+\frac{3}{\pi^2 t^2}\Big(\Delta_\tau+8 \Big) \Big(10 F_4 (v) - 7 F_2(v)^2\Big)+O(e^{-2\pi t})\eea
around the non--separating node. Importantly, the right hand side of \C{step3} contains $\Delta_v F_4 (v)$ which also appears in \C{Z2}, but not in \C{Z3} or \C{Z5}. Thus to \C{step3}, we add $120\mathcal{Z}_2$ such that this contribution cancels. This leads to the asymptotic expansion
\bea \label{step4}&&\Delta \Big(\mathcal{Z}_5 +12\mathcal{Z}_3 - 24 \varphi^2\Big) + 120\mathcal{Z}_2 = -\frac{28\pi^2 t^2}{3} -56 \pi t g(v) -52 F_2 (v) - 84 g(v)^2 \non \\ &&+\frac{3}{\pi t}\Big[ 28 g(v)F_2 (v) -12 \Big(D_3 - D_3^{(1)}(v)\Big) + 48\Big(E_3 - g_3 (v)\Big) -\frac{7}{\pi} \Delta_v F_2(v)^2\Big]\non \\ &&+\frac{3}{\pi^2 t^2}\Big[30 F_4 (v)-7\Big(\Delta_\tau +8\Big)F_2 (v)^2\Big]+O(e^{-2\pi t})\eea
around the non--separating node. Now subtracting $3\mathcal{Z}_5$ from \C{step4} and using \C{Z5}, we see that several terms cancel in the $O(1/t)$ and $O(1/t^2)$ contributions, giving us the asymptotic expansion
\bea \label{step5}&&\Delta \Big(\mathcal{Z}_5 +12\mathcal{Z}_3 - 24 \varphi^2\Big) + 120\mathcal{Z}_2 - 3\mathcal{Z}_5 = -\frac{28\pi^2 t^2}{3} -56 \pi t g(v) -52 F_2 (v) - 84 g(v)^2\non \\ &&-\frac{3}{\pi t} \Big[ 8 g(v) F_2 (v)  +\frac{11}{2\pi} \Delta_v F_2(v)^2\Big]- \frac{3}{\pi^2t^2} \Big[ 65 F_2 (v)^2 + \frac{11}{2} \Delta_\tau F_2(v)^2\Big] + O(e^{-2\pi t}).\eea 
Now the terms involving $\Delta_v F_2(v)^2$ and $\Delta_\tau F_2(v)^2$ in \C{step5} are precisely proportional to \C{prop} and can be accounted for by the asymptotic expansion of $132\mathcal{Z}_3$, leading to 
\bea \label{step6}&&\Delta \Big(\mathcal{Z}_5 +12\mathcal{Z}_3 - 24 \varphi^2\Big) + 120\mathcal{Z}_2 - 3\mathcal{Z}_5 +132 \mathcal{Z}_3= -2\pi^2 t^2 -12 \pi t g(v) \non \\ &&-18 g(v)^2 - 30 F_2 (v) -\frac{90}{\pi t} g(v) F_2 (v) -\frac{225}{2\pi^2 t^2} F_2(v)^2 + O(e^{-2\pi t}).\eea
Strikingly, the right hand side of \C{step6} is the asymptotic expansion of $-72\varphi^2$ around the non--separating node on using \C{exp1}, leading to the Poisson equation
\be \label{finalexp}\Big(\Delta -3\Big)\Big(\mathcal{Z}_5 +12\mathcal{Z}_3 - 24 \varphi^2\Big) = -24\Big(5\mathcal{Z}_2 + 7\mathcal{Z}_3\Big).\ee 
Though we have deduced \C{finalexp} by an analysis of the asymptotic expansions of the various string invariants only around the non--separating node where we have neglected the exponentially suppressed contributions, the manner in which various simplifications occur leading to a compact expression leads us to conjecture that the Poisson equation \C{finalexp} is satisfied all over the moduli space of genus two Riemann surfaces. Apart from proving or disproving this statement, it will be interesting to try to obtain more Poisson equations involving other string invariants by an analysis of their asymptotic expansions around the non--separating node.

\appendix

\section{The expression for $\Delta_\tau D_3^{(1)} (v)$}

In order to calculate the action of $\Delta_\tau$ on the elliptic modular graph $D_3^{(1)} (v)$, it is very useful to perform the analysis by varying the complex structure of the torus to obtain the eigenvalue equation~\cite{Basu:2015ayg,Basu:2016xrt,Basu:2016kli,Kleinschmidt:2017ege,Basu:2019idd}. The relevant variations involving the Green function are given by~\cite{Verlinde:1986kw,DHoker:1988pdl,DHoker:2015gmr,Basu:2015ayg}
\be \label{Beltrami}\p_\mu g(z_1-z_2) = -\frac{1}{\pi}\int_{\S}d^2 z \p_z g(z-z_1) \p_z g(z-z_2),\ee
and
\be \label{Beltrami2}\overline\p_\mu\p_\mu g(z_1-z_2)=0,\ee
which follows from analyzing variations with Beltrami differential $\mu$.

Since the Laplacian $\Delta_\tau$ is given in terms of these variations by
\be \label{relBel}\Delta_\tau = \overline\p_\mu\p_\mu \ee
this enables us to calculate the action of $\Delta_\tau$ on $D_3^{(1)} (v)$. 

Thus using \C{Beltrami2} and \C{relBel}, we get that
\bea \frac{1}{2}\Delta_\tau D_3^{(1)} (v) = \int_{\S} \frac{d^2 z}{2\tau_2} g(z-v) \p_\mu g(z) \overline\p_\mu  g(z) +\Big[ \int_{\S} \frac{d^2 z}{2\tau_2} \p_\mu g(z-v) g(z) \overline\p_\mu g(z)+c.c.\Big],\eea
which we now evaluate using \C{Beltrami}. Each term has two worldsheet holomorphic and two anti--holomorphic derivatives acting on the Green function. Using the single--valuedness of the Green function in \C{Green}, we now integrate by parts and remove as many derivatives as possible using the relations 
\bea \label{eigen}\overline\p_w\p_z g(z-w) &=& \pi \delta^2 (z-w) - \frac{\pi}{\tau_2}, \non \\
\overline\p_z\p_z g(z-w) &=& -\pi \delta^2 (z-w) + \frac{\pi}{\tau_2}.\eea
This leads to
\bea \label{final}\Delta_\tau D_3^{(1)} (v) = 2 E_3 + 4 g_3 (v) -\frac{2\tau_2}{\pi} \p_v g_2 (v) \overline\p_v g_2 (v).\eea
Rewriting the last term in \C{final} using the relation
\be \tau_2\p_v g_2 (v) \overline\p_v g_2(v) = \frac{1}{8}\Delta_v F_2(v)^2 - \pi g(v) F_2 (v)\ee
leads to \C{acttau}.


\begin{thebibliography}{10}

\bibitem{DHoker:2015gmr}
E.~D'Hoker, M.~B. Green, and P.~Vanhove, ``{On the modular structure of the
  genus-one Type II superstring low energy expansion},'' {\em JHEP} {\bf 08}
  (2015) 041,
\href{http://www.arXiv.org/abs/1502.06698}{{\tt 1502.06698}}.

\bibitem{DHoker:2015wxz}
E.~D'Hoker, M.~B. Green, O.~Gurdogan, and P.~Vanhove, ``{Modular Graph
  Functions},'' {\em Commun. Num. Theor. Phys.} {\bf 11} (2017) 165--218,
\href{http://www.arXiv.org/abs/1512.06779}{{\tt 1512.06779}}.

\bibitem{Green:1999pv}
M.~B. Green and P.~Vanhove, ``{The Low-energy expansion of the one loop type II
  superstring amplitude},'' {\em Phys.Rev.} {\bf D61} (2000) 104011,
\href{http://www.arXiv.org/abs/hep-th/9910056}{{\tt hep-th/9910056}}.

\bibitem{Green:2008uj}
M.~B. Green, J.~G. Russo, and P.~Vanhove, ``{Low energy expansion of the
  four-particle genus-one amplitude in type II superstring theory},'' {\em
  JHEP} {\bf 0802} (2008) 020,
\href{http://www.arXiv.org/abs/0801.0322}{{\tt 0801.0322}}.

\bibitem{Richards:2008jg}
D.~M. Richards, ``{The One-Loop Five-Graviton Amplitude and the Effective
  Action},'' {\em JHEP} {\bf 0810} (2008) 042,
\href{http://www.arXiv.org/abs/0807.2421}{{\tt 0807.2421}}.

\bibitem{Green:2013bza}
M.~B. Green, C.~R. Mafra, and O.~Schlotterer, ``{Multiparticle one-loop
  amplitudes and S-duality in closed superstring theory},'' {\em JHEP} {\bf 10}
  (2013) 188,
\href{http://www.arXiv.org/abs/1307.3534}{{\tt 1307.3534}}.

\bibitem{Basu:2015ayg}
A.~Basu, ``{Poisson equation for the Mercedes diagram in string theory at genus
  one},'' {\em Class. Quant. Grav.} {\bf 33} (2016), no.~5, 055005,
\href{http://www.arXiv.org/abs/1511.07455}{{\tt 1511.07455}}.

\bibitem{Zerbini:2015rss}
F.~Zerbini, ``{Single-valued multiple zeta values in genus 1 superstring
  amplitudes},'' {\em Commun. Num. Theor. Phys.} {\bf 10} (2016) 703--737,
\href{http://www.arXiv.org/abs/1512.05689}{{\tt 1512.05689}}.

\bibitem{DHoker:2016mwo}
E.~D'Hoker and M.~B. Green, ``{Identities between Modular Graph Forms},'' {\em
  J. Number Theor.} {\bf 189} (2018) 25--88,
\href{http://www.arXiv.org/abs/1603.00839}{{\tt 1603.00839}}.

\bibitem{Basu:2016xrt}
A.~Basu, ``{Poisson equation for the three loop ladder diagram in string theory
  at genus one},'' {\em Int. J. Mod. Phys.} {\bf A31} (2016), no.~32, 1650169,
\href{http://www.arXiv.org/abs/1606.02203}{{\tt 1606.02203}}.

\bibitem{Basu:2016kli}
A.~Basu, ``{Proving relations between modular graph functions},'' {\em Class.
  Quant. Grav.} {\bf 33} (2016), no.~23, 235011,
\href{http://www.arXiv.org/abs/1606.07084}{{\tt 1606.07084}}.

\bibitem{Basu:2016mmk}
A.~Basu, ``{Simplifying the one loop five graviton amplitude in type IIB string
  theory},'' {\em Int. J. Mod. Phys.} {\bf A32} (2017), no.~14, 1750074,
\href{http://www.arXiv.org/abs/1608.02056}{{\tt 1608.02056}}.

\bibitem{DHoker:2016quv}
E.~D'Hoker and J.~Kaidi, ``{Hierarchy of Modular Graph Identities},'' {\em
  JHEP} {\bf 11} (2016) 051,
\href{http://www.arXiv.org/abs/1608.04393}{{\tt 1608.04393}}.

\bibitem{Kleinschmidt:2017ege}
A.~Kleinschmidt and V.~Verschinin, ``{Tetrahedral modular graph functions},''
  {\em JHEP} {\bf 09} (2017) 155,
\href{http://www.arXiv.org/abs/1706.01889}{{\tt 1706.01889}}.

\bibitem{DHoker:2019blr}
E.~D'Hoker and M.~B. Green, ``{Exploring transcendentality in superstring
  amplitudes},'' {\em JHEP} {\bf 07} (2019) 149,
\href{http://www.arXiv.org/abs/1906.01652}{{\tt 1906.01652}}.

\bibitem{Basu:2019idd}
A.~Basu, ``{Eigenvalue equation for the modular graph $C_{a,b,c,d}$},'' {\em
  JHEP} {\bf 07} (2019) 126,
\href{http://www.arXiv.org/abs/1906.02674}{{\tt 1906.02674}}.

\bibitem{Gerken:2019cxz}
J.~E. Gerken, A.~Kleinschmidt, and O.~Schlotterer, ``{All-order differential
  equations for one-loop closed-string integrals and modular graph forms},''
  {\em JHEP} {\bf 01} (2020) 064,
\href{http://www.arXiv.org/abs/1911.03476}{{\tt 1911.03476}}.

\bibitem{Gerken:2020yii}
J.~E. Gerken, A.~Kleinschmidt, and O.~Schlotterer, ``{Generating series of all
  modular graph forms from iterated Eisenstein integrals},'' {\em JHEP} {\bf
  07} (2020), no.~07, 190, \href{http://www.arXiv.org/abs/2004.05156}{{\tt
  2004.05156}}.

\bibitem{Gerken:2020aju}
J.~E. Gerken, ``{Basis Decompositions and a Mathematica Package for Modular
  Graph Forms},'' \href{http://www.arXiv.org/abs/2007.05476}{{\tt 2007.05476}}.

\bibitem{Gerken:2020xfv}
J.~E. Gerken, A.~Kleinschmidt, C.~R. Mafra, O.~Schlotterer, and B.~Verbeek,
  ``{Towards closed strings as single-valued open strings at genus one},''
  \href{http://www.arXiv.org/abs/2010.10558}{{\tt 2010.10558}}.

\bibitem{DHoker:2005vch}
E.~D'Hoker and D.~H. Phong, ``{Two-loop superstrings VI: Non-renormalization
  theorems and the 4-point function},'' {\em Nucl. Phys.} {\bf B715} (2005)
  3--90,
\href{http://www.arXiv.org/abs/hep-th/0501197}{{\tt hep-th/0501197}}.

\bibitem{DHoker:2005jhf}
E.~D'Hoker, M.~Gutperle, and D.~H. Phong, ``{Two-loop superstrings and
  S-duality},'' {\em Nucl. Phys.} {\bf B722} (2005) 81--118,
\href{http://www.arXiv.org/abs/hep-th/0503180}{{\tt hep-th/0503180}}.

\bibitem{Berkovits:2005df}
N.~Berkovits, ``{Super-Poincare covariant two-loop superstring amplitudes},''
  {\em JHEP} {\bf 01} (2006) 005,
\href{http://www.arXiv.org/abs/hep-th/0503197}{{\tt hep-th/0503197}}.

\bibitem{Berkovits:2005ng}
N.~Berkovits and C.~R. Mafra, ``{Equivalence of two-loop superstring amplitudes
  in the pure spinor and RNS formalisms},'' {\em Phys. Rev. Lett.} {\bf 96}
  (2006) 011602,
\href{http://www.arXiv.org/abs/hep-th/0509234}{{\tt hep-th/0509234}}.

\bibitem{DHoker:2013fcx}
E.~D'Hoker and M.~B. Green, ``{Zhang-Kawazumi Invariants and Superstring
  Amplitudes},'' {\em Journal of Number Theory} {\bf 144} (2014) 111,
\href{http://www.arXiv.org/abs/1308.4597}{{\tt 1308.4597}}.

\bibitem{DHoker:2014oxd}
E.~D'Hoker, M.~B. Green, B.~Pioline, and R.~Russo, ``{Matching the $D^{6}R^{4}$
  interaction at two-loops},'' {\em JHEP} {\bf 01} (2015) 031,
\href{http://www.arXiv.org/abs/1405.6226}{{\tt 1405.6226}}.

\bibitem{Pioline:2015qha}
B.~Pioline, ``{A Theta lift representation for the Kawazumi-Zhang and Faltings
  invariants of genus-two Riemann surfaces},'' {\em J. Number Theor.} {\bf 163}
  (2016) 520--541, \href{http://www.arXiv.org/abs/1504.04182}{{\tt
  1504.04182}}.

\bibitem{DHoker:2017pvk}
E.~D'Hoker, M.~B. Green, and B.~Pioline, ``{Higher genus modular graph
  functions, string invariants, and their exact asymptotics},'' {\em Commun.
  Math. Phys.} {\bf 366} (2019), no.~3, 927--979,
\href{http://www.arXiv.org/abs/1712.06135}{{\tt 1712.06135}}.

\bibitem{DHoker:2018mys}
E.~D'Hoker, M.~B. Green, and B.~Pioline, ``{Asymptotics of the $D^8
  \mathcal{R}^4$ genus-two string invariant},'' {\em Commun. Num. Theor. Phys.}
  {\bf 13} (2019) 351--462, \href{http://www.arXiv.org/abs/1806.02691}{{\tt
  1806.02691}}.

\bibitem{Basu:2018bde}
A.~Basu, ``{Eigenvalue equation for genus two modular graphs},'' {\em JHEP}
  {\bf 02} (2019) 046,
\href{http://www.arXiv.org/abs/1812.00389}{{\tt 1812.00389}}.

\bibitem{DHoker:2020prr}
E.~D'Hoker, C.~R. Mafra, B.~Pioline, and O.~Schlotterer, ``{Two-loop
  superstring five-point amplitudes. Part I. Construction via chiral splitting
  and pure spinors},'' {\em JHEP} {\bf 08} (2020) 135,
  \href{http://www.arXiv.org/abs/2006.05270}{{\tt 2006.05270}}.

\bibitem{DHoker:2020tcq}
E.~D'Hoker, C.~R. Mafra, B.~Pioline, and O.~Schlotterer, ``{Two-loop
  superstring five-point amplitudes II: Low energy expansion and S-duality},''
  \href{http://www.arXiv.org/abs/2008.08687}{{\tt 2008.08687}}.

\bibitem{Basu:2020pey}
A.~Basu, ``{Poisson equations for elliptic modular graph functions},''
  \href{http://www.arXiv.org/abs/2009.02221}{{\tt 2009.02221}}.

\bibitem{DHoker:2020uid}
E.~D'Hoker and O.~Schlotterer, ``{Identities among higher genus modular graph
  tensors},'' \href{http://www.arXiv.org/abs/2010.00924}{{\tt 2010.00924}}.

\bibitem{Basu:2020iok}
A.~Basu, ``{Relations between elliptic modular graphs},'' {\em JHEP} {\bf 12}
  (2020) 195, \href{http://www.arXiv.org/abs/2010.08331}{{\tt 2010.08331}}.

\bibitem{DHoker:2020aex}
E.~D'Hoker, A.~Kleinschmidt, and O.~Schlotterer, ``{Elliptic modular graph
  forms I: Identities and generating series},''
  \href{http://www.arXiv.org/abs/2012.09198}{{\tt 2012.09198}}.

\bibitem{Basu:2020goe}
A.~Basu, ``{Integrating simple genus two string invariants over moduli
  space},'' \href{http://www.arXiv.org/abs/2012.14006}{{\tt 2012.14006}}.

\bibitem{Kawazumi}
N.~Kawazumi, ``{Johnson's Homomorphisms and the Arakelov Green Function},''
  \href{http://www.arXiv.org/abs/0801.4218}{{\tt 0801.4218}}.

\bibitem{Zhang}
S.~W. Zhang, ``{Gross--Schoen Cycles and Dualising Sheaves},'' {\em Invent.
  Math.} {\bf 179 (1)} (2010) 1--73,
  \href{http://www.arXiv.org/abs/0812.0371}{{\tt 0812.0371}}.

\bibitem{Moore:1986rh}
G.~W. Moore, ``{Modular Forms and Two Loop String Physics},'' {\em Phys. Lett.}
  {\bf B176} (1986)
369.

\bibitem{Wentworth}
R.~Wentworth, ``{The Asymptotics of the Arakelov--Green's Function and
  Faltings' Delta Invariant},'' {\em Commun. Math. Phys.} {\bf 137} (1991)
  427--459.

\bibitem{Jong}
R.~De~Jong, ``{Asymptotic Behavior of the Kawazumi--Zhang Invariant for
  Degenerating Riemann Surfaces},'' {\em Asian J. Math.} {\bf 18 (3)} (2014)
  507--524, \href{http://www.arXiv.org/abs/1207.2353}{{\tt 1207.2353}}.

\bibitem{Verlinde:1986kw}
E.~P. Verlinde and H.~L. Verlinde, ``{Chiral Bosonization, Determinants and the
  String Partition Function},'' {\em Nucl. Phys.} {\bf B288} (1987)
357.

\bibitem{DHoker:1988pdl}
E.~D'Hoker and D.~H. Phong, ``{The Geometry of String Perturbation Theory},''
  {\em Rev. Mod. Phys.} {\bf 60} (1988)
917.

\end{thebibliography}

\providecommand{\href}[2]{#2}\begingroup\raggedright\endgroup

\end{document}